\documentclass[aps,prl,groupedaddress,twocolumn]{revtex4}
\usepackage{graphicx,psfrag,bm,amsmath,amsfonts,amssymb,color,subfigure}
\newcommand{\be}{\begin{equation}}
\newcommand{\ee}{\end{equation}}
\newcommand{\beq}{\begin{eqnarray}}
\newcommand{\eeq}{\end{eqnarray}}
\newcommand{\ba}{\begin{array}}
\newcommand{\ea}{\end{array}}
\newcommand{\bea}{\begin{eqnarray}}
\newcommand{\eea}{\end{eqnarray}}
\newcommand{\ex}[1]{\mbox{e}^{#1}}

\newcommand{\eref}[1]{(\ref{#1})}

\newcommand{\mcal}[1]{\mathcal{#1}}

\newcommand{\g}{\gamma_{\perp}}
\newcommand{\gp}{\gamma_{\|}}
\newcommand{\om}{\omega}
\newcommand{\Om}{\Omega}

\newcommand{\bma}{\begin{matrix}}
\newcommand{\ema}{\end{matrix}}

\newcommand{\Omu}{\Omega_{\mu}}
\newcommand{\psimu}{\Psi_\mu(\bm{x})}

\newcommand{\bx}{\bm{x}}
\newcommand{\bxp}{\bm{x}^{\prime}}

\begin{document}
\title{Theory of the spatial structure of non-linear lasing modes}
\author{Hakan E. T\"ureci}
\affiliation{Institute of Quantum Electronics, ETH Zurich, 8093
Zurich, Switzerland}
\author{A. Douglas Stone and Li Ge}
\affiliation{Department of Applied Physics, P. O. Box 208284, Yale University,
New Haven, CT 06520-8284, USA}

\date{\today}

\begin{abstract}
A self-consistent integral equation is formulated and solved iteratively which determines the steady-state lasing modes of open multi-mode lasers. These modes are naturally decomposed in terms of frequency dependent biorthogonal modes of a linear wave equation and not in terms of resonances of the cold cavity. A one-dimensional cavity laser is analyzed and the lasing mode is found to have non-trivial spatial structure even in the single-mode limit. In the multi-mode regime spatial hole-burning and mode competition is treated exactly. The formalism generalizes to complex, chaotic and random laser media.
\end{abstract}
\maketitle

The steady-state electric field within and outside of a single or
multi-mode laser arises as a solution of the non-linear coupled
matter-field equations, the simplest of which are the two-level
Maxwell-Bloch equations treated below.  While the basic equations
involved have been known for many years, and many aspects of their
temporal dynamics have been studied \cite{mandel_book_cnlo}, relatively little
progress has been made in understanding the spatial structure of the
non-linear electric field, particularly in the case of multi-mode
solutions for which spatial hole-burning and other non-linear effects
are critical.  It is natural to attempt to understand
the non-linear solutions in terms of solutions of a linear wave
equation.  The two standard choices are either the hermitian
solutions of a perfectly reflecting (closed) passive laser cavity
\cite{haken_light2}, or the non-hermitian non-orthogonal resonances
of the open passive cavity \cite{LeungLY94,ChingLvSTY98}).  In fact
the intuitive picture of a lasing mode is that it arises when one of
the resonances of the passive cavity is ``pulled" up to the real axis
by adding gain to the resonator.  Often comparison of the numerically
generated lasing
modes with calculated linear resonances do show strong similarities
in spatial structure, providing useful interpretation of lasing
modes \cite{HarayamaDI03,HarayamaSI05}, although not a predictive
theory.  However with the current interest in complex laser cavities
based on wave-chaotic shapes \cite{Tureci05,SchwefelTureci04},
photonic bandgap media \cite{PainterLSYODK99,RyuKPHLK02} or random media
\cite{Cao03,Cao05} it is important to have a quantitative and predictive theory of the lasing
states, as the numerical simulations required to solve the time-dependent
Maxwell-Bloch equations are time-consuming and not easy to
interpret.

In recent work we have formulated a theory of
steady-state multi-mode lasing which addresses these concerns
\cite{hsc06}.  The theory implies that the natural linear basis for
decomposing lasing solutions is the dual set of biorthogonal states
corresponding to constant outgoing and incoming Poynting vector at
infinity at the lasing frequencies (referred to as ``constant flux"
(CF) states).  Our theory shows that even in conventional lasers it
is incorrect to regard the lasing modes as corresponding to a single
resonance of the passive cavity and that multiple spatial frequencies
occur even when there is a single lasing frequency close to the
frequency of a single passive cavity resonance. These multiple
spatial frequencies arise because several CF states contribute to a
single lasing mode.  Note that biorthogonal modes have been used
extensively in resonator theory \cite{siegman_book} (notably for the
case of unstable resonators) but have not previously been applied to
multimode lasing theory.  For multimode lasing the main difficulty is
treating modal interactions and the related effects of spatial
hole-burning \cite{haken63}. We sketch below an efficient method for
treating these effects exactly which can in principle be used in
designing laser cavities to predict power output and
tailor the mode spectrum of the laser.  The techniques
are illustrated for the simple case of a one-dimensional
edge-emitting laser.

We begin with the semiclassical laser equations within the rotating wave and
slowly-varying envelope approximations (see Ref.~\cite{hsc06} for a derivation) 
\bea
\dot{e}  &=& \frac{i}{2\om_a} \left[ \om_a^2  + \frac{c^2}{n^2(\bm{x})}
\nabla^2 \right] e + \frac{2i\pi\om_a}{n^2(\bx)} p
\label{eqMB0}\\
\dot{p} &=& -\g p + \frac{g^2}{i\hbar} eD  \label{eqMB1}\\
\dot{D} &=& \gp\left( D_0-D \right) - \frac{2}{i\hbar}\left( e p^* -
p e^* \right) \label{eqMB2 }
\eea
describing a laser comprised of a uniform gain medium of two-level atoms (with level spacing $\hbar \omega_a$) embedded in a background dielectric medium/cavity with arbitrary spatially varying index of refraction $n(\bx)$.  Here $e(\bx,t)$, $p(\bx,t)$ are the envelopes of the field and polarization, $D(\bx,t)$ is the inversion, $D_0$ is the pump strength, $g$ is the dipole matrix element, $\gamma_{\perp}$ and $\gp$ are damping constants for $p$, $D$. As usual the fast variation of the
fields at $\omega_a$ is removed and the actual fields are given by $(E,P)=(e,p)\,\ex{i\om_a t} + c.c.$ For simplicity we take $e$, $p$ to be scalar fields, appropriate for the 1D case with TM polarization that we will discuss below; the same scalar form would apply for planar random or chaotic cavities.  $n(\bx)$ is the (possibly) spatially dependent index of refraction of the cavity.

We assume a steady-state lasing solution
which is multi-periodic in time: $e(\bm{x},t) = \sum_\mu \Psi_\mu(\bm{x})
\ex{-i\Omega_\mu t}$, $p(\bm{x},t) = \sum_\mu p_\mu(\bm{x})
\ex{-i\Omega_\mu t}$. In contrast to standard modal expansions, not only the lasing
frequencies $\{ \Omega_\mu \}$
but the spatial mode functions $\{
\Psi_\mu(\bm{x}) \} $ are assumed to be unknown.  

\begin{figure}[hbt]
\includegraphics[clip,width=\linewidth]{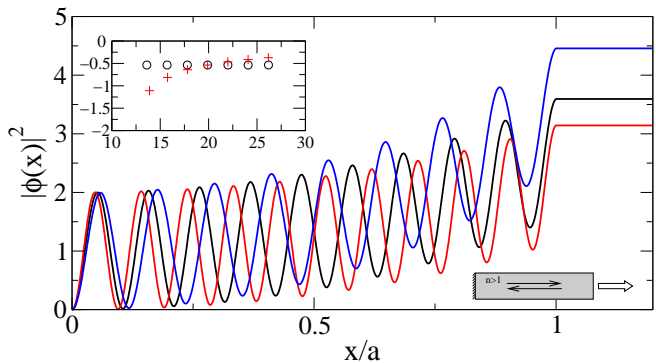}
\caption{The spatial structure of three consecutive CF modes
calculated at the external frequency $k=19.90$ and $n_0=1.5$. The CF
frequencies are $k_m = 19.90  - 0.536\,i$ (black), $21.99 - 0.466\,i$
(red), and $17.81 - 0.640\,i$ (blue). The upper inset
compares the CF mode frequencies ($+$) with the quasibound mode
frequencies ($\circ$) in the range $m=[6,12]$. Lower inset:
schematic of the edge-emitting laser cavity.}
\label{figcfstates}
\end{figure}

Such a multi-periodic solution requires that the inversion is approximately
stationary \cite{FuH91}, implying \cite{hsc06} that
each lasing mode must satisfy the self-consistent equation.
\begin{equation}
\psimu =  \frac{i\frac{4\pi \omega_a^2 g^2}{\hbar c^2}D_0}{  -i \Omu
+\gamma_{\perp}}
\int_{\mcal{D}} d\bxp \frac{ G(\bx,\bxp | \Om_\mu) \Psi_\mu (\bxp)}{1 +
\sum_\nu g(\Om_\nu) |\Psi_\nu (\bxp) |^2}
\label{eqscint}
\end{equation}
where $g(\Omega_\mu)$ is the gain profile evaluated at the lasing
frequency, $\mcal{D}$ is the cavity domain and $G$ is the Green function of the cavity wave
equation $[ \nabla^2  + n^2(\bx) k^2 ]\,
G(\bm{x},\bm{x}'|\om) = \delta^3(\bm{x}-\bm{x}')$
with purely outgoing boundary conditions and $k =
\omega/c$ is an external wavevector of the
lasing solution at infinity (for multimode solutions $k = k_\mu = \Omega_\mu/c$).
Henceforth we set $c=1$ and use wavevector and frequency interchangeably.
With  these non-hermitian boundary conditions the spectral
representation $G(\bx,\bxp | k)$ is of the form:
\begin{equation}
G(\bx,\bxp |k) =  \sum_m
\frac{\varphi_{m}(\bx,k)\bar{\varphi}^*_{m}(\bxp,k)}{ 2 \eta_m \, k_a
(k -k_m)}.
\label{eqspecrep2}
\end{equation}
Here the functions $\varphi_{m}(\bx,k)$ are the CF states which
satisfy $-\nabla^2 \varphi_m(\bm{x},k) = n^2(\bx) k_m ^2
\varphi_m(\bm{x},k)$ with the non-hermitian boundary condition of only
outgoing waves of wavevector $k$ at the cavity boundary. For the
special case of a 1D
cavity of length $a$ considered below (see Fig.~\ref{figcfstates},
inset) this condition is just $\partial_x
\varphi_m (x)|_a = +ik \varphi_m (a)$.  Note this differs subtly but
importantly from the quasi-bound state boundary
condition where the complex eigenvalue $k_m$ replaces the real
wavevector $k$ \cite{hsc06}.
The dual set of functions
$\bar{\varphi}_{m}(x,k)$ satisfy the complex conjugate
differential equation with the boundary conditions $\partial_x
\bar{\varphi}_m (x)|_a = -ik \bar{\varphi}_m (a)$
corresponding to
constant incoming flux. In general these functions satisfy the biorthogonality
relations  \cite{morse&feshbach}:  $\int_{\mcal{D}} \, d\bx
\, \bar{\varphi}^*_m(\bx,k)  n^2(\bx) \varphi_n(\bx,k) = \eta_m
(k) \delta_{mn}$,
and are also complete.  These relations make it
possible to expand an arbitrary lasing solution
\begin{equation}
\Psi_\mu (\bx) = \sum a_m^\mu \varphi_m^{\mu} (\bx)
\label{eqmpansatz}
\end{equation}
so that each $\Psi_\mu$ is a vector in the
space of biorthogonal functions. Here, $\varphi^\mu_m (\bx) =
\varphi_m(\bx, k_\mu)$ and in
what follows we define $k^{\mu}_m=k_m(k_\mu)$. By substitution of \eref{eqmpansatz}
into Eq.~\eref{eqscint} and use of the biorthogonality relations one finds:
\begin{eqnarray}
a^\mu_m = \frac{i D_0 \g}{(\g - i
k_\mu)}\frac{1}{\eta_m^\mu (k_\mu - k_m^{\mu})}  &  \nonumber \\
\times \int_{\mcal{D}}
\frac{ d\bx \, \bar{\varphi}_m^{\mu*} (\bx) \sum_p a^\mu_p
\varphi^\mu_p(\bx)
}{1 +  \sum_{\nu, q,r} g(k_\nu) a^\nu_q a_r^{\nu*} \varphi^\nu_q(\bx)
\varphi_r^{\nu*} (\bx)} &
\label{eqam}
\end{eqnarray}
where we have rescaled the pump $2 \pi k_a g^2 D_0/\hbar \g  \to D_0$, and
measured
electric field in units $e_c = \hbar \sqrt{\gp \g
}/2g$.

Equation (\ref{eqam}) is the key result of our work; it
determines the lasing mode(s), each of which is a superposition of CF
states which depends on its lasing frequency, $k_\mu$, and the pump
power, $D_0$.
It is useful to regard
Eq. (\ref{eqam}) as defining a map of the complex vector space of
coefficients $ {\bf a^\mu} = (a^\mu_1,
a^\mu_2,\ldots)$ into itself,
where the actual non-zero lasing solution is a fixed point of this
map.  Above the lasing threshold for each mode, $D^\mu_{0t}$, we find
that the non-zero solutions are stable fixed points and trial vectors
flow to them under iteration of the map while the trivial zero
solutions, which below $D^\mu_{0t}$ are stable, become unstable.
Note that the map is proportional to $[k_\mu - k_m]^{-1}$, favoring
the CF state with complex wavevector close to the real lasing wavevector,
and it is also proportional to $[ \g - i k_\mu]^{-1}$,
insuring that the lasing frequency is near the center of the gain
profile.  It can be shown that in the high-finesse limit in
which the imaginary part of the CF frequency is very much smaller
than the real spacing between them only one CF state dominates the
lasing state (the "single-pole approximation"), and this CF state is
virtually identical to the corresponding linear resonance
\cite{hsc06}.  In this limit the picture of a single resonance being
``pulled" up to the real axis is valid.  However in many realistic
cases this limit is not realized and the actual lasing solution is
the superposition of CF states determined by Eq. (\ref{eqam}).  

Eq. (\ref{eqam}) determines the lasing frequencies
as well.  For the first lasing mode and a uniform index resonator
this is particularly simple.  At threshold the CF states in the
denominator of the
integrand can be ignored and biorthogonality leads to a simple
relation:
\be
a_m = \frac{i D_0 \g/n^2}{(\g - i
k)}\frac{1}{(k-k_m(k))} \, a_m.
\ee
For a non-trivial solution we must
have $a_m \neq 0$ for some $m$ and hence the coefficient must be real
and equal to unity.  The reality condition determines that the possible lasing
frequencies at threshold are $k_t^{(m)}= \frac{\g q_m (k_t^{(m)})}{\g + \kappa_m (k_t^{(m)})}$
where $k_m  \equiv q_m (k) -i\kappa_m (k)$ (we
suppress the index $\mu$ here). Furthermore,
the modulus unity
condition determines the threshold pumping:
\be
D_{0t}^{(m)}
= n^2 \kappa_m \left[ 1 + \frac{q_m^2}{(\g +  \kappa_m)^2} \right].
\label{eqfirstthr}
\ee
The CF state $m$ and associated frequency leading
to the lowest
threshold will be the first lasing mode.
Note that in contrast to the
traditional mode-pulling formula \cite{lamb_book} where the cavity
mode frequency is a fixed value, here the single-mode laser frequency
is determined by the solution of a self-consistent equation
(a
transcendental equation for the 1D case). Nonetheless for high
finesse cavities this condition agrees with standard results: the
lasing frequency is very
close to the cavity resonance nearest to the gain center, pulled
towards the gain center by an amount which depends on the relative
magnitude of
$\g$ vs. $\kappa_m$ (which is approximately the
resonance linewidth) \cite{hsc06}.

Above threshold the lasing mode is found by
initially choosing the lasing wavevector, $k=k_t^{(m)}$, calculating
the sets $\{\varphi_m \},\{\bar{\varphi}_m \}$ corresponding to that
choice and then iterating Eq. (\ref{eqam}) starting from a trial vector
$\bm{a}(0)$ to yield output $\bm{a}(1)$.  A natural choice for the
initial vector is $a_m = 1, a_p = 0, \forall_p \neq m$, where $m$ is
the dominant component at threshold, calculated from the above
relations.  As noted, the ``lasing map" has the property that below
the lasing threshold, $D_{0t}$, the iterated vector, $\bf{a} \to 0$,
and above this threshold it converges to a finite value which defines
the spatial structure of the lasing mode in terms of the CF states.
There is one crucial addition necessary to complete the algorithm.
Note that Eq. (\ref{eqam}) is invariant under multiplication of the vector
$\bm{a}$ by a global phase $e^{i\theta}$, so iteration of (\ref{eqam}) can
never determine a unique non-zero solution.  Therefore it is
necessary to fix the ``gauge" of the solution by demanding that we
solve (\ref{eqam}) with the constraint of a certain global phase (typically
we take the dominant $a_m$ to be real).  Thus after each iteration of
(\ref{eqam}) we must
adjust the lasing frequency to restore the phase of $a_m$; it is just
this gauge fixing requirement which causes the lasing frequency to
flow from our initial guess to the correct value above
threshold. The invariance of (\ref{eqam}) under global phase changes
guarantees that the
frequency thus found is independent of the particular gauge choice.
For multi-mode lasing we repeat this procedure for each vector
$\bm{a}^\mu$.  In the single-mode regime only one of these vectors will
flow to a non-zero fixed point.
\begin{figure}[hbt]
\includegraphics[clip,width=\linewidth]{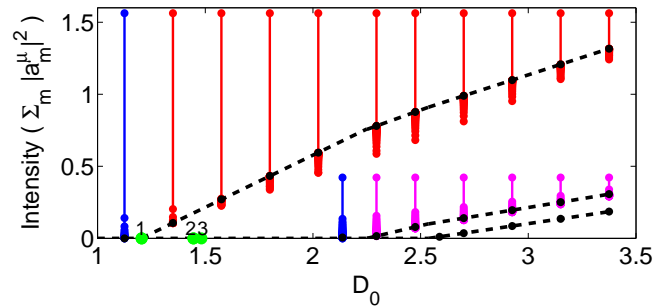}
\caption{Convergence and solution of the multimode lasing map for 1D
resonator with $n_0 = 1.5$, $k_a = 19.89$ and
$\gamma_{\perp} = 4.0$ vs. pump $D_0$.  Three modes lase
in this range.  At threshold they correspond to CF states $m=8,9,10$ with
threshold lasing frequencies $k_t^{(8)}=18.08, k_t^{(9)}=19.91$, and
$k_t^{(10)}=21.76$, and non-interacting thresholds
$D_0^{(9)} = 1.204, D_0^{(10)}=1.445, D_0^{(8)}= 1.482$
(green dots). $k_t^{(9)} \approx k_a$ and $m=9$ has the lowest
threshold.  Due to mode competition, modes 2,3 do not lase until much
higher values ($D_0 = 2.25,2.53$).  Each mode is
represented by an 11 component vector of CF states; we plot the
sum of $|a_m|^2$ vs. pump $D_0$. Below threshold
the vectors flows to zero (blue dots).  For $D_0 \geq 1.204$
the sum flows (red dots) to a non-zero value (black dashed line), and
above $D_0 = 2.25,2.53$, two additional non-zero vector fixed points (modes)
are found (convergence only shown for modes 1,2).}
\label{figmmlaser}
\end{figure}
This behavior of the multi-mode
lasing map is illustrated in Fig.~\ref{figmmlaser} for the simple
uniform index 1D
cavity corresponding to an edge-emitting laser with a perfect mirror
at the origin and an index step at $x=a$ (inset to
Fig.~\ref{figcfstates}).  Below the
first threshold, determined from Eq.~\ref{eqfirstthr} above, the entire set of
vectors $\bm{a}^\mu$ flow to zero; above that threshold the first
lasing mode turns on and its intensity grows linearly with pump
strength.  Due to its non-linear interaction with other modes, the
turn-on of the second and third lasing modes is dramatically
suppressed, leading to a factor of four increase in the interval of
single-mode operation.  The intensity shows slope discontinuities at
higher thresholds as seen in normal laser operation. 
Note that in 
this approach effects of spatial hole-burning
and mode competition are treated exactly and not in the
near-threshold approximation (cubic non-linearity) traditionally used
\cite{haken_light2, lamb_book} which greatly {\it underestimates} the
output power \cite{hsc06}.

\begin{figure}[hbt]
\includegraphics[clip,width=\linewidth]{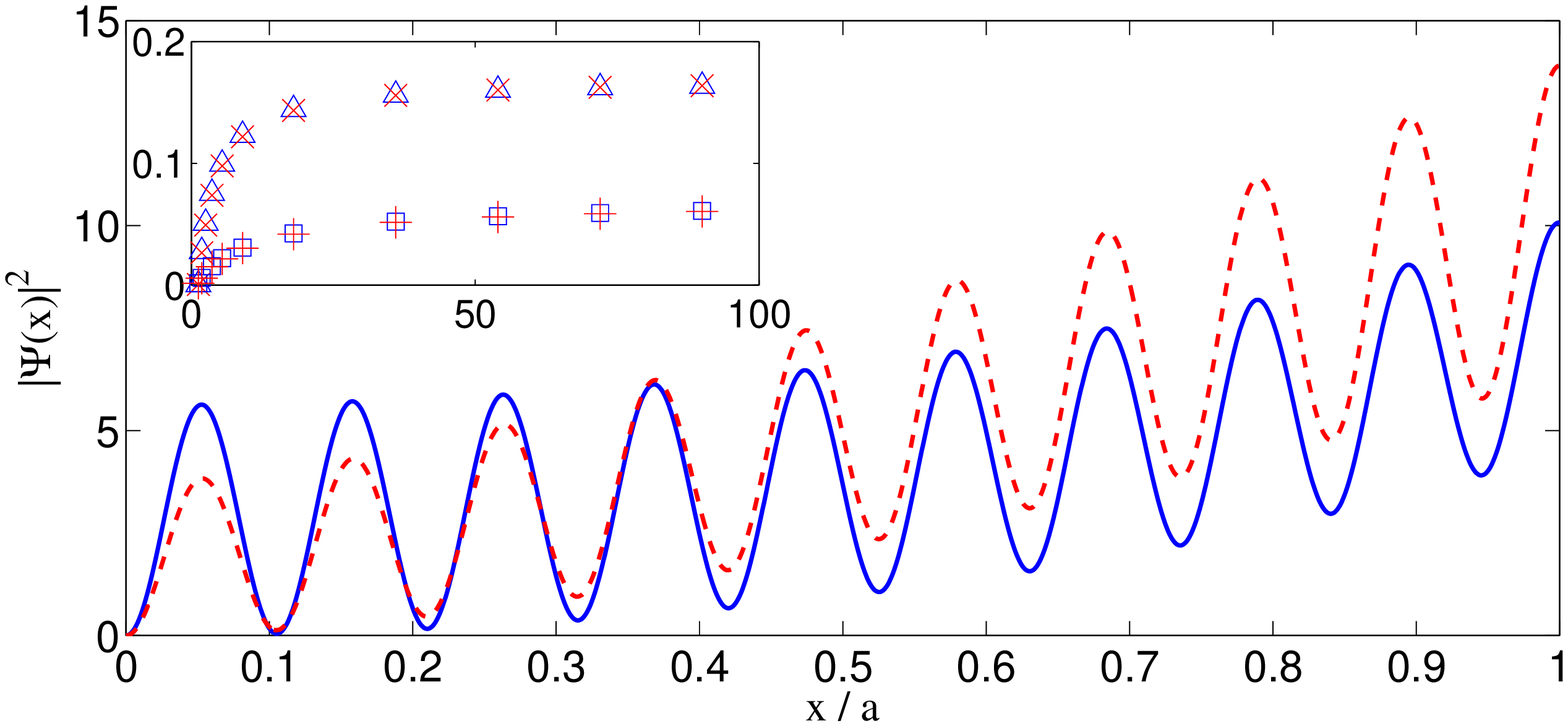}
\caption{Non-linear electric field intensity for a single-mode
edge-emitting laser with $n_0=1.5, k_a = 19.89,\gamma_\perp=0.5,
D_0=9$.  The full field (red line) has an
appreciably larger amplitude at the output $x=a$ than the
``single-pole" approximation (blue) which neglects the sideband CF
components.
Inset: The ratio of the two largest CF sideband
components to that of the central pole for $n_0=1.5$ ($\square$,
$\times$) and $n_0=3$ ($\square$,$+$) vs.  pump strength $D_0$.}
\label{figsidebands}
\end{figure}

We now consider the spatial structure of the CF
states defining the lasing modes.
The linear resonances (quasi-bound
states) of this system are easily found \cite{hsc06}, they are
complex sine waves of wavevector $n_0 k_m^{qb}a  = \pi (m + 1/2) - i
/2 \ln [(n_0 +1)/(n_0 -1)]$.  The constant linewidth $Im[k_m^{(qb)}] 
\equiv - \kappa_m^{(qb)}$
follows from the Fresnel transmissivity of 
dielectric boundary at normal incidence.  Note
that 
$\kappa_m^{(qb)} >0$ here, corresponding to amplification with 
increasing $x$.
From Eq.~\eref{eqmpansatz} we expect the lasing mode to involve
several CF states and differ most from a single cavity resonance for a low finesse cavity; thus
we consider relatively small index, $n_0 = 1.5$.  The CF states 
depend on the lasing
wavevector, $k$.  In Fig. ~(\ref{figcfstates}) we choose it to be the
real part of the wavevector of the
$9^{th}$ resonance, $ k^{qb}_{9}$, and plot the 7 closest CF eigenvalues,
$k_m$ (inset, Fig.~\ref{figcfstates}).  The $9^{th}$ CF state has
$k_{9} \approx k^{qb}_{9}$ and within the cavity is very close to the 
$m=9$
resonance
\cite{hsc06}, but  the other $k_m$, while they have $Re[k_m] \approx 
Re[k_m^{(qb)}]$
(hence similar FSR) have substantially larger or smaller $\kappa_m$ 
than $\kappa_m^{(qb)}$.
Hence only the $m=9$ CF state is close to a 
linear resonance, emphasizing that CF states
are {\it not} resonances.
We plot several of these states in Fig.~\ref{figcfstates}, showing 
their different
amplification rates.
The actual lasing mode will be 
the sum of several of
these CF modes with different spatial frequencies and amplification
rates. In Fig. (\ref{figsidebands}) we plot such a mode.
Standard modal expansions in laser theory are equivalent to choosing
only the central CF state and missing the contribution of these
spatial ``side-bands" \cite{hsc06}. The inset to
Fig.~\ref{figsidebands} shows that near threshold only one CF state
dominates (one can show that the other components are of order the
cube of the dominant component).  But well above threshold the two
nearest neighbor CF states are each $15\%$ of the main component and
since one of these has higher amplification rate, the final effect is
to increase the output power by more than $43\%$ (see
Fig.~\ref{figsidebands}).  The
sidebands are still $6\%$ of the dominant component when the index is
increased to $n_0=3$, leading to an increase in output power by
$26\%$ (see inset, Fig.~\ref{figsidebands}).
The formalism we have presented here is
ideal for treating random or wave-chaotic lasers for which the output
directionality, output power and mode spectrum are very hard to
predict with heuristic arguments.  Moreover in these systems
the finesse typically is parametrically smaller than unity, suggesting
that each lasing mode will consist of many CF states
without a dominant component.  We have considered here the borderline
case of a cavity with finesse of order unity, so there is a dominant
CF component, but the spatial sidebands are still appreciable.  Since
our method treats the non-linearity and mode-competition exactly we
anticipate that it may be useful in designing efficient semiconductor
microcavity lasers.
\begin{acknowledgments}
This work was supported by NSF grant DMR 0408636 and by the Aspen Center for Physics.
\end{acknowledgments}


\end{document}